\documentclass[12pt,psamsfonts,reqno]{amsart}

\theoremstyle{definition}

\theoremstyle{remark}

\newcommand{\vev}[1]{\left\langle #1 \right\rangle}
\newcommand{\ket}[1]{\left |  #1 \right \rangle}
\newcommand{\bra}[1]{\left \langle  #1 \right |}

\newcommand {\CalN} {\mathcal N}

\newcommand {\CalX} {\mathcal X}

\newcommand {\BI}   {\mathbb I}

\newcommand {\BR}   {\mathbb R}
\newcommand {\BZ}   {\mathbb Z}
\newcommand {\BC}   {\mathbb C}

\newcommand{\bV}{\mathbf{V}}

\newcommand{\bH}{\mathbf{H}}
\newcommand{\bM}{\mathbf{M}}

\newcommand{\bX}{\mathbf{X}}

\newcommand{\bY}{\mathbf{Y}}

\newcommand{\msF}{\mathscr{F}}

\newcommand{\msS}{\mathscr{S}}
\newcommand{\msW}{\mathscr{W}}
\newcommand{\si}{\mathsf{i}}
\newcommand{\sn}{\mathsf{n}}

\newcommand{\sk}{\mathsf{k}}

\newcommand{\sA}{\mathsf{A}}

\newcommand{\sP}{\mathsf{P}}
\newcommand{\sS}{\mathsf{S}}
\newcommand{\sV}{\mathsf{V}}
\newcommand{\sT}{\mathsf{T}}

\newcommand{\sY}{\mathsf{Y}}

\newcommand{\g}{\mathfrak{g}}

\newcommand{\fq}{\mathfrak{q}}

\DeclareMathOperator{\rk} {rk}

\DeclareMathOperator{\Li}{Li}

\renewcommand{\Im}{\operatorname{Im}}

\newcommand{\U}{\mathrm{U}}

\numberwithin{equation}{section}

\usepackage[margin=1in]{geometry}

\usepackage{amsmath}
\usepackage{amssymb}
\usepackage{amsxtra}
\usepackage{amscd}
\usepackage[mathscr]{euscript}

\usepackage[pagebackref=false]{hyperref}
\hypersetup{colorlinks = false, ,extension = notused,linkcolor = blue, anchorcolor = red,citecolor = blue,filecolor = red,pagecolor = red,urlcolor = blue}
\usepackage[numbers,sort&compress]{natbib}
\usepackage{hypernat}

\usepackage{iftex}
\ifxetex
        \usepackage{fontspec}
\else
\fi

\usepackage{float}
\usepackage{tikz-cd}
\usepackage[mathscr]{euscript}

\begin{document}

\title{Twisted reduction of quiver W-algebras}

\author{Taro Kimura}
\author{Vasily Pestun}

\address{Taro Kimura, Keio University, Japan}
\address{Vasily Pestun, IHES, France} 

\begin{abstract} 
 We consider the $k$-twisted Nekrasov--Shatashvili limit (NS$_k$ limit) of 5d (K-theoretic) and 6d (elliptic) quiver gauge theory, where one of the multiplicative equivariant parameters is taken to be the $k$-th root of unity.
 We obtain the extended center of the associated $q$-deformed quiver W-algebras constructed by our formalism~\cite{Kimura:2015rgi,Kimura:2016dys,Kimura:2017hez}, which provides gauge theoretic proof of Bouwknegt--Pilch's statement on the relation to the representation ring of quantum affine algebra.
\end{abstract}

\maketitle 

\tableofcontents

\parskip=4pt

\section{Introduction}

The geometric approach to 4d $\CalN = 2$ (5d $\CalN = 1$) gauge theory, initiated by Seiberg--Witten~\cite{Seiberg:1994rs,Seiberg:1994aj}, gives rise to an interesting correspondence to the algebraic integrable system~\cite{Gorsky:1995zq,Martinec:1995by,Donagi:1995cf,Gorsky:1996hs}.
This correspondence allows a quantum deformation using the equivariant ($\Omega$-background) parameters $(\epsilon_1, \epsilon_2)$, introduced to write down the gauge theory partition function $Z(a;\epsilon_1,\epsilon_2;\fq)$, depending on the Coulomb moduli $(a_i)$ and the coupling constant $\fq$~\cite{Moore:1997dj,Nekrasov:2002qd}.
Seiberg--Witten's prepotential is reproduced from the partition function in the {\em classical limit} $\epsilon_{1,2} \to 0$~\cite{Nekrasov:2003rj,Nakajima:2003pg,Nakajima:2005fg,Braverman:2004vv,Braverman:2004cr}:
\begin{align}
 \msF(a;\fq) =
 \lim_{\epsilon_{1,2} \to 0} \epsilon_1 \epsilon_2
 \log Z(a;\epsilon_1,\epsilon_2;\fq)
 \, .
 \label{eq:prepot}
\end{align}
It has been then pointed out by Nekrasov--Shatashvili~\cite{Nekrasov:2009rc} that the partition function shows a similar scaling behavior in the limit $\epsilon_2 \to 0$ while keeping $\epsilon_1$ finite, which is so-called Nekrasov--Shatashvili (NS) limit:
\begin{align}
 \msW(a;\epsilon_1;\fq) = \lim_{\epsilon_{2} \to 0}
 \epsilon_2 \log Z(a;\epsilon_1,\epsilon_2;\fq)
 \label{eq:twistpot}
\end{align}
where the twisted superpotential of the effective 2d theory $\msW(a;\epsilon_1;\fq)$ plays a central role in the correspondence to the quantum integrable system.
The crucial idea is to identify the superpotential $\msW(a;\epsilon_1;\fq)$ as the Yang--Yang function of the integrable system~\cite{Yang:1968rm}.
Under this identification, the SUSY vacuum given by the twisted $F$-term condition is determined by the Bethe ansatz equation of the quantum integrable system since the derivative of Yang--Yang function with the quasimomenta gives rise to it.

Quantization of the correspondence between gauge theory and integrable system is pursued by considering the situation with generic equivariant parameters.
Although, in this case, we do not expect a scaling behavior for the partition function, like \eqref{eq:prepot} and \eqref{eq:twistpot}, it has an interpretation as the conformal block of 2d conformal field theory (CFT), which is known as the AGT relation~\cite{Alday:2009aq,Wyllard:2009hg} and its $q$-analogue for 5d $\CalN=1$ gauge theory~\cite{Awata:2009ur}.
The underlying conformal symmetry is W-algebra W$(G)$ which shows dependence on the gauge group $G$ when the gauge group $G$ is simple.
On the other hand, we have established another connection with W-algebra, dubbed quiver W-algebra W$(\Gamma)$, whose conformal symmetry is determined by the quiver structure of gauge theory $\Gamma$~\cite{Kimura:2015rgi,Kimura:2016dys,Kimura:2017hez} when the gauge group $G = \prod_{i \in {\Gamma_0}} SU(n_i)$ 
and the hypermultiplets are in the fundamental and bifundamental representation.
The  construction of $\mathrm{W}(\Gamma)$-algebra has a close connection with the representation theory of quiver $\Gamma$:
the generating current of quiver W-algebra is given by the $qq$-character~\cite{Nekrasov:2015wsu,Nekrasov:2016qym,Nekrasov:2016ydq,Nekrasov:2017rqy,Nekrasov:2017gzb} of the fundamental representation associated to the quiver.
This is a natural generalization of the classical and NS limit of gauge theory, where the classical and quantum Seiberg--Witten curves are realized using the character and $q$-character of the fundamental representations~\cite{Nekrasov:2012xe,Nekrasov:2013xda}.
The $q$-character, introduced by Frenkel--Reshetikhin~\cite{Frenkel:1998} and by Knight~\cite{Knight:1995JA} for the affine Yangian, describes the representation ring of quantum affine algebra.
This is consistent with the geometric $q$-Langlands correspondence since, in the NS limit, the W-algebra is reduced to the classical commutative Poisson algebra, which is isomorphic to quantum affine algebra at the critical level.
This {\em classical} relation is promoted to quantum geometric $q$-Langlands correspondence between the $q$-deformed W-algebra and quantum affine algebra away from the classical limit and critical level~\cite{Aganagic:2017smx}.
This is indeed consistent with the quiver gauge theory construction of W-algebras under generic equivariant parameters $(\epsilon_1,\epsilon_2)$.

It is known that 5d gauge theory compactified on a circle realizes the K-theory lift, corresponding to the relativistic integrable system~\cite{Nekrasov:1996cz}, where the additive equivariant parameters are replaced with multiplicative ones, $(q_1,q_2)$, as defined in \eqref{eq:omega_mult}.
Then the classical and NS limit corresponds to $q_{1,2} \to 1$ and $q_2 \to 1$, respectively.
For such a multiplicative deformation parameter, one can consider another {\em classical} limit $q_{1,2} \to \zeta_k$ with the $k$-th root of unity $\zeta_k = \exp \left( 2 \pi \iota / k \right)$, and $\iota = \sqrt{-1}$.
The root of unity limit of 5d gauge theory has been considered~\cite{Kimura:2011zf,Kimura:2011gq} to explore the instanton counting on $A$-type ALE space, since the $\BZ_k$ orbifold projection is performed in this limit.
See also~\cite{Itoyama:2013mca,Itoyama:2014pca,Yoshioka:2015voz,Itoyama:2017rvf}.
From this point of view, it is natural to ask what is the geometric representation theoretical meaning of such a twisted limit for quiver variety.
In this paper we consider $k$-twisted NS limit, $q_2 \to \zeta_k$ while keeping $q_1$ finite, which we call the NS$_k$ limit, and elucidate its role in quiver W-algebra to address the question above.
We remark that Bouwknegt--Pilch~\cite{Bouwknegt1997qru,Bouwknegt:1998da} have studied the limit corresponding to the NS$_k$ limit in the context of the $q$-deformed W-algebra.
Our approach in this paper would provide gauge theory perspective for their statement.

The remaining part of this paper is organized as follows.
In order to fix our notation, in Sec.~\ref{sec:gauge-th}, we begin with the anatomy of quiver gauge theory partition function in 5d and 6d.
We perform the saddle point analysis of the gauge theory partition function in Sec.~\ref{sec:saddle} to clarify the asymptotic behavior in the NS and NS$_k$ limit.
We obtain an expression of the saddle point equation in terms of the $\sY$-function, and construct the $q_k$-character generated by the iWeyl$_k$ reflection in the NS$_k$ limit of 5d and 6d theory.
We discuss the operator formalism of quiver gauge theory in Sec.~\ref{sec:op} and reproduce the $q_k$-character in the NS$_k$ limit, which generates the extended center elements of quiver W-algebra.
In Sec.~\ref{sec:discussion} we address several concluding remarks and discussions.
Appendix~\ref{sec:asymptotic} summarizes the formulae used to analyze the asymptotic limit of gauge theory partition function.

\subsection*{Acknowledgements}
The work of T.K. was supported in part by JSPS Grant-in-Aid for Scientific Research (No.~JP17K18090), the MEXT-Supported Program for the Strategic Research Foundation at Private Universities ``Topological Science'' (No.~S1511006), and JSPS Grant-in-Aid for Scientific Research on Innovative Areas ``Topological Materials Science'' (No.~JP15H05855), and ``Discrete Geometric Analysis for Materials Design'' (No.~JP17H06462).
 This project has received funding from the European Research Council (ERC) under the European Union's Horizon 2020 research and innovation program (QUASIFT grant agreement 677368).

\section{Quiver gauge theory}\label{sec:gauge-th}

The quiver gauge theory has multiple vector and (bifundamental) hypermultiplets characterized by the quiver $\Gamma$, which consists of nodes (vertices) $i \in \Gamma_0$ and edges (arrows) $e \in \Gamma_1$.
The quiver $\Gamma$ defines a matrix $(c)_{ij}$, called the quiver Cartan matrix,
\begin{align}
 c_{ij} & = 2 \delta_{ij}
 - \#(e: i \to j) - \#(e: j \to i) 
 \, .
 \label{eq:Cartan_quiver}
\end{align}
We assign $\U(\sn_i)$ gauge group to $i$-th node,%
\footnote{%
We can also utilize the supergroup $\U(\sn_i^+|\sn_i^-)$ for gauge symmetry.
See~\cite{Kimura:2019msw} for details.
}
and multiplicative mass parameter $\mu_{e} = e^{m_e}$ to the bifundamental matter on the edge $e$.
In this paper, we use the 5d (K-theoretic; multiplicative) notation:
\begin{align}
 (q_1, q_2) = (e^{\epsilon_1}, e^{\epsilon_2})
 \label{eq:omega_mult}
\end{align}
with $q = q_1 q_2$, and assume the matrix $(c)_{ij}$ is symmetric, namely simply-laced quiver.
See~\cite{Kimura:2017hez} for generalization to non-simply-laced quiver.

The partition function of 4d $\CalN=2$ gauge theory is given by integration over the instanton moduli space, which is computable using the localization formula~\cite{Moore:1997dj,Nekrasov:2002qd}.
Furthermore, 5d theory compactified on a circle $\BR^4 \times S^1$ corresponds to the K-theoretic uplift.
The 5d partition function is obtained as the equivariant index functor of the structure sheaf, which converts the additive Chern character to the multiplicative class,
\begin{align}
 \BI \left[ \sum_k x_k \right]
 & =
 \prod_{k} \left( 1 - x_k^{-1} \right)
 \, .
 \label{eq:ind}
\end{align}
This is the Dolbeault convention of the index since it obeys the reflection relation
\begin{align}
 \BI \left[ \bX^\vee \right]
 & =
 (-1)^{\rk \bX} \left( \det \bX \right) \BI \left[ \bX \right]
 \, ,
 \label{eq:ind_ref}
\end{align}
while the Dirac index obeys $\BI \left[ \bX^\vee \right] = (-1)^{\rk \bX} \BI \left[ \bX \right]$.
These two conventions are equivalent when the quiver satisfies the conformal condition $\sum_{j \in \Gamma_0} c_{ij} \sn_j = \sn_i^\text{f} + \sn_i^{\text{af}}$, where $\sn_i^\text{f}$ and $\sn_i^\text{af}$ are the numbers of the fundamental and antifundamental hypermultiplets for the node $i \in \Gamma_0$.
Otherwise, we have to shift the Chern--Simons level.

Similarly the partition function of 6d theory defined on $\BR^4 \times T^2$ is given by the equivariant elliptic genus 
\begin{align}
 \BI_p \left[ \sum_k x_k \right]
 & =
 \prod_{k} \theta(x_k^{-1};p)
 \label{eq:ind_6d}
\end{align}
where the theta function defined
\begin{align}
 \theta(x;p) = (x;p)_\infty (px^{-1};p)_{\infty}
\end{align}
with the $q$-shifted factorial ($q$-Pochhammer symbol)
\begin{align}
 (x;q)_n = \prod_{m=0}^{n-1}( 1 - x q^m ) 
\end{align}
and the multiplicative modulus of the compactified two-torus $p = e^{2\pi \iota \tau}$.
We remark that this index also obeys the reflection relation \eqref{eq:ind_ref}, and the 6d index \eqref{eq:ind_6d} is reduced to the 5d index \eqref{eq:ind} in the limit $p \to 0$ ($\Im \tau \to \infty$).
The 6d theory has to satisfy the conformal condition due to the gauge/modular anomaly, so that we do not have to care about the convention issue.
See~\cite{Kimura:2016dys} for details.

In order to describe the Chern character at a fix point of the moduli space under the torus action labeled by the partition $(\lambda_{i,\alpha,k})$, we define a set
\begin{align}
 x_{i,\alpha,k} = q_2^{\lambda_{i,\alpha,k}} q_1^{k-1} \nu_{i,\alpha}
 \, , \quad
 \CalX_i = \{ x_{i,\alpha,k} \}_{i \in \Gamma_0, \alpha = 1,\ldots,\sn_i,k = 1,\ldots,\infty}
 \, , \quad
 \CalX = 
 \bigsqcup_{i \in \Gamma_0} \CalX_i
 \, .
\end{align}
where $(\nu_{i,\alpha})_{i \in \Gamma_0, \alpha =1,\ldots, \sn_i}$ is the multiplicative Coulomb moduli for $i$-th gauge node.
We also define a set of empty configuration $(\lambda_{i,\alpha}) = \emptyset$,
\begin{align}
 \mathring{x}_{i,\alpha,k} = q_1^{k-1} \nu_{i,\alpha}
 \, , \quad
 \mathring{\CalX}_{i} = \{ \mathring{x}_{i,\alpha,k} \}_{i \in \Gamma_0, \alpha = 1,\ldots,\sn_i,k = 1,\ldots,\infty}
 \, , \quad
 \mathring{\CalX} = 
 \bigsqcup_{i \in \Gamma_0} \mathring\CalX_{i}
 \, .
\end{align}

The Chern character of the observable sheaf evaluated at the fixed point for the node $i \in \Gamma_0$ is given by
\begin{align}
 \bY_i = (1 - q_1) \sum_{x \in \CalX_i} x
 \, ,
 \label{eq:obs-sheaf}
\end{align}
and we denote by $\bY_i^{[n]}$ the $n$-th Adams operation of $\bY_i$, which is interpreted as the UV chiral ring operator of gauge theory,
\begin{align}
 \bY_i^{[n]} = (1 - q_1^n) \sum_{x \in \CalX_i} x^n
 \, .
 \label{eq:obs-sheaves}
\end{align}
Then the vector and bifundamental hypermultiplets contribution to the Chern character is written using the observable sheaf,
\begin{align}
 \bV_i
 & =
 \frac{1}{(1 - q_1)(1 - q_2)} \bY_i^\vee \bY_i
 =
 \frac{1 - q_1^{-1}}{1 - q_2}
 \sum_{(x,x') \in \CalX_i \times \CalX_i} \frac{x'}{x}
 \\
 \bH_{e:i \to j}
 & =
 - \frac{\mu_e}{(1 - q_1)(1 - q_2)} \bY_i^\vee \bY_j
 =
 - \mu_e \frac{1 - q_1^{-1}}{1 - q_2}
 \sum_{(x,x') \in \CalX_i \times \CalX_j} \frac{x'}{x}
\end{align}
The total character has a compact form using the mass-deformed half Cartan matrix
\begin{align}
 c_{ij}^+
 & = \delta_{ij} - \sum_{e:i \to j} \mu_e^{-1}
 \label{eq:Cartan_half1}
\end{align}
as
\begin{align}
 \sum_{i \in \Gamma_0} \bV_i
 + \sum_{e:i \to j} \bH_{e:i \to j}
 & =
 \sum_{(x,x') \in \CalX \times \CalX}
 \left( c_{\si(x)\si(x')}^+\right)^\vee
 \frac{1 - q_1^{-1}}{1 - q_2}
 \frac{x'}{x}
\end{align}
where $\si: \CalX \to \Gamma_0$ is the node label such that $\si(x) = i$ if $x \in \CalX_i$.

The fundamental and antifundamental hypermultiplet contributions are obtained from the bifundamental matter by freezing the auxiliary gauge node, namely replacing $\bY_i \to \bM_i$ and $\bY_i^\vee \to \widetilde{\bM}_i$,
\begin{align}
 \bH_i^\text{f}
 = - \frac{1}{(1 - q_1)(1 - q_2)} \bY^\vee_i \bM_i
 = \frac{q_1^{-1}}{1 - q_2}
 \sum_{(x,x') \in \CalX_i \times \CalX_i^\text{f}} \frac{x'}{x}
 \\[.5em]
 \bH_i^\text{af}
 = - \frac{1}{(1 - q_1)(1 - q_2)} \widetilde{\bM}_i \bY_i
 = - \frac{1}{1 - q_2}
 \sum_{(x,x') \in \CalX_i^\text{af} \times \CalX_i} \frac{x'}{x} 
\end{align}
where 
\begin{align}
 \bM_i
 = \sum_{x \in \CalX_i^\text{f}} x 
 = \sum_{f=1}^{\sn_i^\text{f}} \mu_{i,f}
 \, , \qquad
 \widetilde{\bM}_i
 = \sum_{x \in \CalX_i^\text{af}} x 
 = \sum_{f=1}^{\sn_i^\text{af}} \tilde{\mu}_{i,f}
 \, ,
\end{align}
with sets of the multiplicative fundamental and antifundamental mass parameters denoted by $\CalX_i^\text{f} = \{\mu_{i,f}\}_{i \in \Gamma_0, f \in [1,\ldots,\sn_i^\text{f}]}$ and $\CalX_i^{\text{af}} = \{\tilde{\mu}_{i,f}\}_{i \in \Gamma_0, f \in [1,\ldots,\sn_i^{\text{af}}]}$.

Applying the index \eqref{eq:ind}, we obtain contributions to the full partition function for 5d quiver gauge theory, including both 1-loop and instanton factors,
\begin{subequations}\label{eq:Z_5d}
\begin{align}
 Z_i^\text{vec}
 & =
 \BI \left[ \bV_i \right]
 =
 \prod_{(x,x') \in \CalX_i \times \CalX_i}
 \left(
 q \frac{x}{x'};q_2
 \right)_\infty
 \left(
 q_2 \frac{x}{x'};q_2
 \right)_\infty^{-1}
 \, ,\\
 Z_{e:i \to j}^\text{bf}
 & =
 \BI \left[ \bH_{e:i \to j} \right]
 =
 \prod_{(x,x') \in \CalX_i \times \CalX_j}
 \left(
 \mu_e^{-1} q \frac{x}{x'};q_2
 \right)_\infty^{-1}
 \left(
 \mu_e^{-1} q_2 \frac{x}{x'};q_2
 \right)_\infty
 \, ,\\
 Z_i^\text{f}
 & =
 \BI \left[ \bH_i^\text{f} \right]
 = \prod_{(x,\mu) \in \CalX_i \times \CalX_i^\text{f}}
 \left( q \frac{x}{\mu}; q_2 \right)_\infty^{-1}
 \, ,\\
 Z_i^{\text{af}}
 & =
 \BI \left[ \bH_i^{\text{af}} \right]
 = \prod_{(\mu,x) \in \CalX_i^{\text{af}} \times \CalX_i}
 \left( q_2 \frac{\mu}{x}; q_2 \right)_\infty
 \, .
\end{align}
\end{subequations}
In addition, the topological term which counts the instanton number is defined
\begin{align}
 Z_i^\text{top}
 & =
 \fq_i^{|\lambda_{i}|}
 =
 \exp
 \left(
 \log \fq_i
 \sum_{x \in \CalX_i, \mathring{x} \in \mathring{\CalX}_{i}} \log_{q_2} (x/\mathring{x})
 \right)
 \, ,
 \label{eq:Ztop_5d}
\end{align}
and the Chern--Simons term with the level $\kappa_i \in \BZ$ for $i \in \Gamma_0$ is
\begin{align}
 Z_i^{\text{CS}}
 & =
 \exp
 \left(
 \frac{\kappa_i}{2} \sum_{x \in \CalX_i}
 \left( \log_{q_2}^2 x - \log_{q_2} x \right)
 \right)
 \, .
\end{align}

Similarly the 6d partition function is obtained with the elliptic index \eqref{eq:ind_6d}
\begin{subequations}\label{eq:Z_6d}
\begin{align}
 Z_i^\text{vec}
 & =
 \BI_p \left[ \bV_i \right]
 =
 \prod_{(x,x') \in \CalX_i \times \CalX_i}
 \Gamma \left( q \frac{x}{x'}; p, q_2 \right)^{-1}
 \Gamma \left( q_2 \frac{x}{x'}; p, q_2 \right)
 \, ,\\
 Z_{e:i \to j}^\text{bf}
 & =
 \BI_p \left[ \bH_{e:i \to j} \right]
 =
 \prod_{(x,x') \in \CalX_i \times \CalX_j}
 \Gamma \left( \mu_e^{-1} q \frac{x}{x'}; p, q_2 \right)
 \Gamma \left( \mu_e^{-1} q_2 \frac{x}{x'}; p, q_2 \right)^{-1}
 \, ,\\
 Z_i^\text{f}
 & =
 \BI_p \left[ \bH_i^\text{f} \right]
 = \prod_{(x,\mu) \in \CalX_i \times \CalX_i^\text{f}}
 \Gamma \left( q \frac{x}{\mu}; p, q_2 \right)
 \, ,\\
 Z_i^{\text{af}}
 & =
 \BI \left[ \bH_i^{\text{af}} \right]
 = \prod_{(\mu,x) \in \CalX_i^{\text{af}} \times \CalX_i}
 \Gamma \left( q_2 \frac{\mu}{x}; p, q_2 \right)^{-1}
 \, ,
\end{align}
\end{subequations}
with the same topological term $Z_i^{\text{top}}$.
We used the elliptic gamma function defined
\begin{align}
 \Gamma(z;p,q)
 & =
 \prod_{n,m \ge 0}
 \frac{1 - z^{-1} p^{n+1} q^{n+1}}{1 - z p^n q^n}
 \, .
\end{align}
We remark that some analytic continuation is necessary to obtain the expressions in terms of the elliptic gamma functions in \eqref{eq:Z_6d}, and we don't have the Chern--Simons term in 6d gauge theory.

\section{Saddle point analysis}
\label{sec:saddle}

The gauge theory partition function is given as a partition sum~\cite{Nekrasov:2002qd,Nekrasov:2003rj}
\begin{align}
 Z = \sum_\lambda Z_\lambda
 \, .
\end{align}
In the NS limit $\epsilon_2 \to 0$, the asymptotic behavior of the partition function reads~\cite{Nekrasov:2009rc}
\begin{align}
 Z(\epsilon_1, \epsilon_2)
 \ \stackrel{\epsilon_2 \to 0}{\longrightarrow} \
 \exp \left( \frac{1}{\epsilon_2} \msW(\epsilon_1) + \cdots \right)
 \label{eq:ZW}
\end{align}
with the twisted superpotential $\msW$ of the effective 2d theory, and the corresponding twisted $F$-term condition is given by
\begin{align}
 \exp \left( \frac{\partial \msW}{\partial a_\alpha} \right) = 1
 \, .
 \label{eq:F-term}
\end{align}
Since we now have a parameter taken to be small $\epsilon_2 \to 0$, we can apply the saddle point analysis with respect to the small parameter $\epsilon_2$, such that the critical configuration $\lambda_*$ dominates in the partition function $Z \sim Z_{\lambda_*}$.
The saddle point equation, corresponding to the $F$-term condition \eqref{eq:F-term} is given by
\begin{align}
 \exp \left( \epsilon_2 \frac{\partial}{\partial \log x} \log Z_{\lambda_*} \right) = 1
 \, ,
\end{align}
where $x$ is a dynamical multiplicative variable specified later.
In the following, we consider the NS and NS$_k$ limit of the gauge theory partition function obtained above \eqref{eq:Z_5d}, applying the corresponding saddle point analysis. 

\subsection{NS limit}

Using the formulae summarized in Appendix~\ref{sec:asymptotic_5d}, we obtain the asymptotic behavior of the 5d full partition function \eqref{eq:Z_5d} in the NS limit:
\begin{subequations} 
 \begin{align}
  Z_i^\text{vec}
  & \longrightarrow \
  \exp
  \left(
  \frac{1}{\epsilon_2} \sum_{(x,x') \in \CalX_i \times \CalX_i}
  L\left(\frac{x}{x'};q_1\right)
  \right)
  \, , \\
  Z_{e:i \to j}^\text{bf}
  & \longrightarrow \
  \exp
  \left(
  - \frac{1}{\epsilon_2} \sum_{(x,x') \in \CalX_i \times \CalX_j}
  L\left(\mu_e^{-1}\frac{x}{x'};q_1\right)
  \right)
  \, , \\
  Z_i^\text{f}
  & \longrightarrow \
  \exp
  \left(
   - \frac{1}{\epsilon_2} \sum_{(x,\mu) \in \CalX_i \times \CalX_i^\text{f}}
  \Li_2 \left( q_1 \frac{x}{\mu} \right)
  \right)
  \, , \\
  Z_i^\text{af}
  & \longrightarrow \
  \exp
  \left(
   \frac{1}{\epsilon_2} \sum_{(\mu,x) \in \CalX_i^{\text{af}} \times \CalX_i}
  \Li_2 \left( \frac{\mu}{x} \right)
  \right)    
 \end{align}
\end{subequations}
where $\Li_2(z)$ is the dilogarithm and the function $L(z;q_1)$ is defined in \eqref{eq:L-func}.

In order to write down the saddle point equation, we introduce the $\sY$-functions
\begin{align}
 \sY_{i,x}^+
 =
 \prod_{x' \in \CalX_i}
 \frac{1 - x'/x}{1 - q_1 x'/x}
 \, , \qquad
 \sY_{i,x}^-
 =
 \prod_{x' \in \CalX_i}
 \frac{1 - x/x'}{1 - q_1^{-1} x/x'}
 \, .
\end{align}
The asymptotic behavior of these two $\sY$-functions are given by
\begin{align}
 \sY_{i,x}^+
 \longrightarrow
 \begin{cases}
  \displaystyle
  (-1)^{\sn_i} \nu_i \, x^{-\sn_i}
  & (x \to 0) \\
  1 & (x \to \infty)
 \end{cases}
 \qquad
 \sY_{i,x}^-
 \longrightarrow
  \begin{cases}
   1 & (x \to 0) \\
  \displaystyle
  (-1)^{\sn_i} \nu_i^{-1} \, x^{-\sn_i}
  & (x \to \infty)
  \end{cases}
 \label{eq:Y-asymptotic}
\end{align}
where we define the Coulomb moduli product
\begin{align}
 \nu_i = \prod_{\alpha=1}^{\sn_i} \nu_{i,\alpha}
 \, .
\end{align}
Since two $\sY$-functions have the same poles and zeros with the asymptotics \eqref{eq:Y-asymptotic}, they are related to each other in the following way
\begin{align}
 \sY_{i,x}^+ = (-1)^{\sn_i} \nu_i \, x^{-\sn_i} \sY_{i,x}^-
 \, .
 \label{eq:Y_pm-relation}
\end{align}
Now we show another derivation of the relation \eqref{eq:Y_pm-relation} which is also applicable to 6d theory.
The $\sY$-function has the following combinatorial formula
\begin{align}
 \sY_{i,x}^+
 & =
 \prod_{\alpha=1}^{\sn_i}
 \left(
 \left(1 - \frac{\nu_{i,\alpha}}{x}\right)
 \prod_{(i,j) \in \lambda_{i,\alpha}}
 \msS\left(\frac{q_1^{i-1} q_2^{j-1} \nu_{i,\alpha}}{x}\right)
 \right)
 \, , \\
 \sY_{i,x}^-
 & =
 \prod_{\alpha=1}^{\sn_i}
 \left(
 \left(1 - \frac{x}{\nu_{i,\alpha}}\right)
 \prod_{(i,j) \in \lambda_{i,\alpha}}
 \msS\left(\frac{x}{q_1^{i} q_2^{j} \nu_{i,\alpha}}\right)
 \right)
 \, ,
\end{align}
where we define
\begin{align}
 \msS(x) = \frac{(1 - q_1 x)(1 - q_2 x)}{(1 - x)(1 - q x)}
 \, .
 \label{eq:S-factor}
\end{align}
Then, we obtain \eqref{eq:Y_pm-relation} using the reflection formula
\begin{align}
 \msS(x) = \msS(q^{-1} x^{-1})
 \, .
\end{align}
We remark that the $\sY$-function has the following expansion
\begin{align}
 \sY_{i,x}^\pm
 =
 \exp
 \left(
  - \sum_{n=1}^\infty \frac{x^{\mp n}}{n} \, \bY_{i}^{[\pm n]}
 \right)
\end{align}
where $\bY_{i}^{[n]}$ is the chiral ring operator defined \eqref{eq:obs-sheaves}.
Thus it is interpreted as the generating function of the gauge theory observable.
This interpretation will play a role in the operator formalism discussed in Sec.~\ref{sec:op}

The derivative of the partition function with respect to the variable $x \in \CalX_i$ gives rise to
\begin{subequations}
\begin{align}
 \exp
 \left(
  \epsilon_2 \frac{\partial}{\partial \log x} \log Z_i^{\text{vec}}
 \right)
 & =
 \prod_{x' \in \CalX_i \backslash \{x\}}
 \frac{1 - q_1 x' / x}{1 - x' / x}
 \frac{1 - x / x'}{1 - q_1 x / x'}
 =
 \frac{-1}{\sY_{i,x}^+ \sY_{i,q_1 x}^-}
\end{align}
\begin{align}
 \exp \left(
 \epsilon_2 \frac{\partial}{\partial \log x}
 \log Z_{e:j \to i}^\text{bf}
 \right)
 & =
 \prod_{x' \in \CalX_j}
 \frac{1 - \mu_e^{-1} x' / x}{1 - \mu_e^{-1} q_1 x' / x}
 = \sY_{j,\mu_e x}^+
 \\[.5em]
 \exp \left(
 \epsilon_2 \frac{\partial}{\partial \log x}
 \log Z_{e:i \to j}^\text{bf}
 \right)
 & =
 \prod_{x' \in \CalX_j}
 \frac{1 - \mu_e^{-1} q_1 x / x'}{1 - \mu_e^{-1} x / x'}
 = \sY_{j,\mu_e^{-1} q_1 x}^-
\end{align}
\begin{align}
 \exp \left(
 \epsilon_2 \frac{\partial}{\partial \log x}
 \log Z_{i}^\text{f}
 \right)
 & =
 \prod_{\mu \in \CalX_i^\text{f}}
 \left( 1 - \frac{q_1 x}{\mu} \right)
 =: \sP_{i,q_1 x}^-
 \\[.5em]
 \exp \left(
 \epsilon_2 \frac{\partial}{\partial \log x}
 \log Z_{i}^\text{af}
 \right)
 & =
 \prod_{\tilde{\mu} \in \CalX_i^\text{af}}
 \left( 1 - \frac{\tilde{\mu}}{x} \right)
 =: \tilde{\sP}_{i,x}^+
 \, . 
\end{align}
\end{subequations}
Adding the contributions from the topological term and the Chern--Simons term, the saddle point condition reads
\begin{align}
 1 =
 \exp \left(
 \epsilon_2 \frac{\partial}{\partial \log x}
 \log Z^\text{tot}
 \right) 
 = - \fq_i \, x^{\kappa_i}
 \frac{\sP_{i,q_1 x}^- \tilde{\sP}_{i,x}^+}{\sY_{i,q_1 x}^- \sY^+_{i,x}}
 \prod_{e: j \to i} \sY_{j,\mu_e x}^+
 \prod_{e: i \to j} \sY_{j,\mu_e^{-1} q_1 x}^-
 \, .
\end{align}
In the NS limit, the $\sY$-function has a cut singularity.
The saddle point equation describes the crossing-cut condition~\cite{Nekrasov:2012xe}, and the corresponding iWeyl reflection is given by
\begin{align}
 \sY_{i,q_1 x}^-
 \ \longrightarrow \
 \fq_i \, x^{\kappa_i}
 \frac{\sP_{i,q_1 x}^- \tilde{\sP}_{i,x}^+}{\sY^+_{i,x}}
 \prod_{e: j \to i} \sY_{j,\mu_e x}^+
 \prod_{e: i \to j} \sY_{j,\mu_e^{-1} q_1 x}^- 
 \, .
\end{align}
For example, for $A_1$ quiver which consists of a single gauge node, there exists a single $\sY$-function.
Although the $\sY$-function itself has a cut singularity in the NS limit, a proper combination of $\sY_x$ and $\sY_x^{-1}$ characterized by the iWeyl reflection turns out to be a regular polynomial function
\begin{align}
 \sT_x :=
 \sY^-_x + \fq_i \, q_1^{-\kappa} \, x^{\kappa}
 \frac{\sP^-_x \tilde{\sP}^+_{q_1^{-1} x}}{\sY^+_{q_1^{-1} x}}
 \, .
\end{align}
This is called the $q$-character of the fundamental representation of $A_1$ quiver~\cite{Nekrasov:2013xda}, and reproduces the 5d Seiberg--Witten curve in the classical limit $q_{1,2} \to 1$.
Using the relation \eqref{eq:Y_pm-relation}, it can be written in terms of only either $\sY^+(x)$ or $\sY^-(x)$.
We remark that the $\sY$-function appearing in the $q$-character is evaluated with the critical configuration $\lambda_*$, and should be replaced with its average $\Big< \sY (x) \Big>$ for generic $(q_1,q_2)$ when the saddle point approximation is not available.

For generic quiver, the $q$-character for the $i$-th fundamental representation is generated by the iWeyl reflection applied to the $i$-th $\sY$-function which plays a role as the highest weight,
\begin{align}
 \sT_{i,x} =
 \sY_{i,x}^-
 +
 \fq_i \, q_1^{-\kappa_i} \, x^{\kappa_i}
 \frac{\sP_{i,x}^- \tilde{\sP}_{i,q_1^{-1} x}^+}{\sY^+_{i,q_1^{-1} x}}
 \prod_{e: j \to i} \sY_{j,\mu_e q_1^{-1} x}^+
 \prod_{e: i \to j} \sY_{j,\mu_e^{-1} x}^- 
 + \cdots
 \, .
\end{align}
The iWeyl reflection, namely the saddle point equation, assures that the $q$-character defined in this way is a regular polynomial function in $x$, and the polynomial degree is determined by the asymptotic behavior of the $\sY$-function, essentially the gauge group rank $\sn_i$.

\subsection{NS$_k$ limit}

We consider the NS$_k$ limit by parametrizing $q_2 = \exp(\epsilon_2 + 2\pi \iota/k)$ and taking $\epsilon_2 \to 0$.
Let $\zeta_k$ be the primitive $k$-th root of unity $\zeta_k = \exp \left( 2 \pi \iota / k\right)$.
The asymptotic behavior of the 5d full partition function \eqref{eq:Z_5d} in the NS$_k$ limit is obtained with the formulae in Appendix~\ref{sec:asymptotic_5d}:
\begin{subequations}\label{eq:Z_5d_NSk}
 \begin{align}
  Z_i^\text{vec}
  & \longrightarrow \
  \exp
  \left(
  \frac{1}{k \epsilon_2} \sum_{(x,x') \in \CalX_i \times \CalX_i}
  L_k\left(\frac{x}{x'};q_1\right)
  \right)
  \, , \\
  Z_{e:i \to j}^\text{bf}
  & \longrightarrow \
  \exp
  \left(
  - \frac{1}{k \epsilon_2} \sum_{(x,x') \in \CalX_i \times \CalX_j}
  L_k\left(\mu_e^{-1}\frac{x}{x'};q_1\right)
  \right)
  \, , \\
  Z_i^\text{f}
  & \longrightarrow \
  \exp
  \left(
   - \frac{1}{k^2 \epsilon_2} \sum_{(x,\mu) \in \CalX_i \times \CalX_i^\text{f}}
  \Li_2 \left( \left( q_1 \frac{x}{\mu} \right)^k \right)
  \right)
  \, , \\
  Z_i^\text{af}
  & \longrightarrow \
  \exp
  \left(
   \frac{1}{k^2 \epsilon_2} \sum_{(\mu,x) \in \CalX_i^{\text{af}} \times \CalX_i}
  \Li_2 \left( \left( \frac{\mu}{x} \right)^k \right)
  \right)
  \, ,
 \end{align}
\end{subequations}
where the function $L_k(z;q_1)$ is defined in \eqref{eq:Lk-func}.
This shows that the asymptotic behavior in the NS$_k$ limit is simply obtained by replacing all the multiplicative variables with the degree-$k$ variables, $x \to x^k$.
This replacement gives rise to $\BZ_k$ symmetry in the NS$_k$ limit, under the transformation $x \to \zeta_k x$, which is analogous to the $\BZ_k$-orbifold implementation using the $k$-th root of unity limit of the $\Omega$-background parameters~\cite{Kimura:2011zf,Kimura:2011gq}.

In the NS limit, the asymptotic behavior of the partition function yields the effective twisted superpotential \eqref{eq:ZW}, which is interpreted as the Yang--Yang function under the gauge/Bethe correspondence~\cite{Nekrasov:2009rc}.
The NS$_k$ limit shows a similar asymptotic behavior \eqref{eq:Z_5d_NSk}, but a slightly different one: all the multiplicative variables are replaced with degree-$k$ variables.
The asymptotic behavior of the partition function is summarized in the 5d notation
\begin{align}
 Z(q_1,q_2)
 \ {\longrightarrow} \
 \begin{cases}
  \displaystyle
  \exp \left( \frac{1}{\log q_2} \mathscr{W}(x;q_1) \right)
  & (\text{NS}: q_2 \to 1) \\[1em]
  \displaystyle
  \exp \left( \frac{1}{k \log q_2^k} \mathscr{W}(x^k; q_1^k) \right)
  & (\text{NS}_k: q_2 \to \zeta_k) \\  
 \end{cases}
 \label{eq:ZW_5d}
\end{align}

We then consider the saddle point equation in the NS$_k$ limit.
In this case we introduce the degree-$k$ ($\BZ_k$-invariant) $\sY$-function,
\begin{align}
 \sY_{i,x}^{(k)+}
 =
 \prod_{x' \in \CalX_i}
 \frac{1 - x'^k/x^k}{1 - q_1^k x'^k/x^k}
 \, , \qquad
 \sY_{i,x}^{(k)-}
 =
 \prod_{x' \in \CalX_i}
 \frac{1 - x^k/x'^k}{1 - q_1^{-k} x^k/x'^k}
 \, ,
 \label{eq:Yk-functions}
\end{align}
which has an alternative expression in terms of the original $\sY$-function,
\begin{align}
 \sY_{i,x}^{(k)\pm}
 & =
 \prod_{r=0}^{k-1} \sY_{i,\zeta_k^r x}^{\pm}
\end{align}
with the similar relation as before,
\begin{align}
 \sY_{i,x}^{(k)+} = \left( (-1)^{\sn_i} \nu_i \, x^{-\sn_i} \right)^k
 \sY_{i,x}^{(k)-}
 \, .
 \label{eq:Yk_pm-relation}
\end{align}
Since the $\sY^{(k)}$-function has the expansion
\begin{align}
 \sY_{i,x}^{(k)\pm}
 & =
 \exp
 \left(
  - k \sum_{n=1}^\infty \frac{x^{\mp kn}}{kn} \bY_i^{[\pm kn]}
 \right)
 \, ,
\end{align}
it turns out to be the generating function, focusing only on $\BZ_k$-invariant gauge theory observables, $\bY_i^{[n]}$ for $n \in k \BZ$.
This implies that the NS$_k$ limit plays a role as the projection into the $\BZ_k$-invariant sector.

The derivative of the partition function in the NS$_k$ limit with respect to the variable $x \in \CalX_i$ gives rise to
\begin{subequations}
\begin{align}
 \exp
 \left(
  k \epsilon_2 \frac{\partial}{\partial \log x} \log Z_i^{\text{vec}}
 \right)
 & =
 \frac{-1}{\sY_{i,x}^{(k)+} \sY_{i,q_1 x}^{(k)-}}
\end{align}
\begin{align}
 \exp \left(
 k \epsilon_2 \frac{\partial}{\partial \log x}
 \log Z_{e:j \to i}^\text{bf}
 \right)
 & =
 \sY_{j,\mu_e x}^{(k)+}
 \\[.5em]
 \exp \left(
 k \epsilon_2 \frac{\partial}{\partial \log x}
 \log Z_{e:i \to j}^\text{bf}
 \right)
 & =
 \sY_{j,\mu_e^{-1} q_1 x}^{(k)-}
\end{align}
\begin{align}
 \exp \left(
 k \epsilon_2 \frac{\partial}{\partial \log x}
 \log Z_{i}^\text{f}
 \right)
 & =
 \prod_{\mu \in \CalX_i^\text{f}}
 \left( 1 - \frac{q_1^k x^k}{\mu^k} \right)
 =: \sP_{i,q_1 x}^{(k)-}
 \\[.5em]
 \exp \left(
 k \epsilon_2 \frac{\partial}{\partial \log x}
 \log Z_{i}^\text{af}
 \right)
 & =
 \prod_{\tilde{\mu} \in \CalX_i^\text{af}}
 \left( 1 - \frac{\tilde{\mu}^k}{x^k} \right)
 =: \tilde{\sP}_{i,x}^{(k)+}
 \, . 
\end{align}
\end{subequations}
Since the topological term and the Chern--Simons term behave as before, the saddle point equation for the NS$_k$ limit is given by
\begin{align}
 1 & =
 \exp \left(
 k \epsilon_2 \frac{\partial}{\partial \log x}
 \log Z^\text{tot}
 \right)
 =
 - \fq_i^k \, x^{k \kappa_i}
 \frac{\sP_{i,q_1 x}^{(k)-} \tilde{\sP}_{i,x}^{(k)+}}
      {\sY_{i,q_1 x}^{(k)-} \sY^{(k)+}_{i,x}}
 \prod_{e: j \to i} \sY_{j,\mu_e x}^{(k)+}
 \prod_{e: i \to j} \sY_{j,\mu_e^{-1} q_1 x}^{(k)-}
 \, .
\end{align}
This leads to the corresponding iWeyl reflection, which we call the iWeyl$_k$ reflection,
\begin{align}
 \sY_{i,q_1 x}^{(k)-}
 \ \longrightarrow \
 \fq_i^k \, x^{k \kappa_i}
 \frac{\sP_{i,q_1 x}^{(k)-} \tilde{\sP}_{i,x}^{(k)+}}
      {\sY^{(k)+}_{i,x}}
 \prod_{e: j \to i} \sY_{j,\mu_e x}^{(k)+}
 \prod_{e: i \to j} \sY_{j,\mu_e^{-1} q_1 x}^{(k)-}
 \, ,
\end{align}
and thus the degree-$k$ $q$-character, the $q_k$-character, is generated by this iWeyl$_k$ reflection
\begin{align}
 \sT^{(k)}_{i,x}
 =
 \sY_{i,x}^{(k)-}
 +
 \fq_i^k \, q_1^{-k \kappa_i} \, x^{k \kappa_i}
 \frac{\sP_{i,x}^{(k)-} \tilde{\sP}_{i,q_1^{-1} x}^{(k)+}}
      {\sY^{(k)+}_{i,q_1^{-1} x}}
 \prod_{e: j \to i} \sY_{j,\mu_e q_1^{-1} x}^{(k)+}
 \prod_{e: i \to j} \sY_{j,\mu_e^{-1} x}^{(k)-}
 + \cdots
 \, .
 \label{eq:q-ch_k}
\end{align}
This is a regular polynomial function in $x^k$, since all the $\sY$-functions and matter factors are functions of $x^k$.
For example, the $q_k$-character for $A_1$ quiver is given by
\begin{align}
 \sT_x^{(k)}
 =
 \sY^{(k)-}_x
 +
 \fq_i^k \, q_1^{-k \kappa} \, x^{k \kappa}
 \frac{\sP^{(k)-}_x \tilde{\sP}^{(k)+}_{q_1^{-1} x}}
      {\sY^{(k)+}_{q_1^{-1} x}}
 \, ,
 \label{eq:qq-ch_k_A1}
\end{align}
which turns out to be a degree-$\sn$ polynomial in $x^k$ for $U(\sn)$ gauge theory, namely the coefficient of $x^n$ for $n \not\in k \BZ$ is zero.
This implies that the corresponding Coulomb branch is parameterized only by the observables $\bY_i^{[n]}$ for $n \in k \BZ$, which is consistent with the $\sY^{(k)}$-function as a generating function of $\BZ_k$ observables $(\bY_i^{[n]})_{i \in \Gamma_0, n \in k\BZ}$.

\subsection{6d theory}

We can apply almost the same analysis to 6d quiver gauge theory.
Using the formulae in Appendix~\ref{sec:asymptotic_6d}, the asymptotic behavior in the NS$_k$ limit is given by
\begin{subequations}\label{eq:Z_6d_NSk}
 \begin{align}
  Z_i^\text{vec}
  & \longrightarrow \
  \exp
  \left(
  \frac{1}{k \epsilon_2} \sum_{(x,x') \in \CalX_i \times \CalX_i}
  L_k\left(\frac{x}{x'};q_1;p\right)
  \right)
  \, , \\
  Z_{e:i \to j}^\text{bf}
  & \longrightarrow \
  \exp
  \left(
  - \frac{1}{k \epsilon_2} \sum_{(x,x') \in \CalX_i \times \CalX_j}
  L_k\left(\mu_e^{-1}\frac{x}{x'};q_1;p\right)
  \right)
  \, , \\
  Z_i^\text{f}
  & \longrightarrow \
  \exp
  \left(
   - \frac{1}{k^2 \epsilon_2} \sum_{(x,\mu) \in \CalX_i \times \CalX_i^\text{f}}
  \Li_2 \left( \left( q_1 \frac{x}{\mu} \right)^k; p \right)
  \right)
  \, , \\
  Z_i^\text{af}
  & \longrightarrow \
  \exp
  \left(
   \frac{1}{k^2 \epsilon_2} \sum_{(\mu,x) \in \CalX_i^{\text{af}} \times \CalX_i}
  \Li_2 \left( \left( \frac{\mu}{x} \right)^k; p \right)
  \right)
  \, .
 \end{align}
\end{subequations}
Then the saddle point equation is obtained using the elliptic $\sY$-function,
\begin{align}
 \sY_{i,x}^+ =
 \prod_{x' \in \CalX_i}
 \frac{\theta(x'/x;p)}{\theta(q_1 x'/x;p)}
 \, , \qquad
 \sY_{i,x}^- =
 \prod_{x' \in \CalX_i}
 \frac{\theta(x/x';p)}{\theta(q_1^{-1} x/x';p)}
 \, ,
\end{align}
and the degree-$k$ function
\begin{align}
 \sY_{i,x}^{(k)+}
 & =
 \prod_{r=0}^{k-1}
 \sY_{i,\zeta_k^r x}^{(k)+}
 =
 \prod_{x' \in \CalX_i}
 \frac{\theta(x'^k/x^k;p^k)}{\theta(q_1^k x'^k/x^k;p^k)}
 \, , \\
 \sY_{i,x}^{(k)-}
 & =
 \prod_{r=0}^{k-1}
 \sY_{i,\zeta_k^r x}^{(k)-}
 =
 \prod_{x' \in \CalX_i}
 \frac{\theta(x^k/x'^k;p^k)}{\theta(q_1^{-k} x^k/x'^k;p^k)}
 \, .
\end{align}
We remark that the reflection relation \eqref{eq:Y_pm-relation} holds also in the elliptic case.
The $\sY^{(k)}$-function has the expansion
\begin{align}
 \sY_{i,x}^{(k)\pm}
 =
 \exp
 \left(
  - k \sum_{m \neq 0} \frac{x^{\mp km}}{km} \frac{\bY_i^{[\pm km]}}{1 - p^{km}}
 \right)
 \, .
\end{align}
Thus it plays a role as a generating function of the $\BZ_k$ observables $(\bY_i^{[n]})_{n \in k \BZ}$ as before.

The derivative of the partition function with the variable $x \in \CalX_i$ is given by
\begin{subequations}
\begin{align}
 \exp
 \left(
  k \epsilon_2 \frac{\partial}{\partial \log x} \log Z_i^{\text{vec}}
 \right)
 & =
 \frac{-1}{\sY_{i,x}^{(k)+} \sY_{i,q_1 x}^{(k)-}}
\end{align}
\begin{align}
 \exp \left(
 k \epsilon_2 \frac{\partial}{\partial \log x}
 \log Z_{e:j \to i}^\text{bf}
 \right)
 & =
 \sY_{j,\mu_e x}^{(k)+}
 \\[.5em]
 \exp \left(
 k \epsilon_2 \frac{\partial}{\partial \log x}
 \log Z_{e:i \to j}^\text{bf}
 \right)
 & =
 \sY_{j,\mu_e^{-1} q_1 x}^{(k)-}
\end{align}
\begin{align}
 \exp \left(
 k \epsilon_2 \frac{\partial}{\partial \log x}
 \log Z_{i}^\text{f}
 \right)
 & =
 \prod_{\mu \in \CalX_i^\text{f}}
 \theta\left(\frac{q_1^k x^k}{\mu^k};p^k\right)
 =: \sP_{i,q_1 x}^{(k)-}
 \\[.5em]
 \exp \left(
 k \epsilon_2 \frac{\partial}{\partial \log x}
 \log Z_{i}^\text{af}
 \right)
 & =
 \prod_{\tilde{\mu} \in \CalX_i^\text{af}}
 \theta\left(\frac{\tilde{\mu}^k}{x^k};p^k\right)
 =: \tilde{\sP}_{i,x}^{(k)+}
 \, . 
\end{align}
\end{subequations}
In 6d gauge theory, there is no Chern--Simons term.
Thus we obtain the saddle point equation for the NS$_k$ limit
\begin{align}
 1 & =
 \exp \left(
 k \epsilon_2 \frac{\partial}{\partial \log x}
 \log Z^\text{tot}
 \right)
 =
 - \fq_i^k \, 
 \frac{\sP_{i,q_1 x}^{(k)-} \tilde{\sP}_{i,x}^{(k)+}}
      {\sY_{i,q_1 x}^{(k)-} \sY^{(k)+}_{i,x}}
 \prod_{e: j \to i} \sY_{j,\mu_e x}^{(k)+}
 \prod_{e: i \to j} \sY_{j,\mu_e^{-1} q_1 x}^{(k)-}
 \, .
\end{align}
We remark that the RHS is modular invariant, up to the coupling shift, due to the conformal condition for the matter content.
The iWeyl$_k$ reflection is given by
\begin{align}
 \sY_{i,q_1 x}^{(k)-}
 \ \longrightarrow \
 \fq_i^k \, 
 \frac{\sP_{i,q_1 x}^{(k)-} \tilde{\sP}_{i,x}^{(k)+}}
      {\sY^{(k)+}_{i,x}}
 \prod_{e: j \to i} \sY_{j,\mu_e x}^{(k)+}
 \prod_{e: i \to j} \sY_{j,\mu_e^{-1} q_1 x}^{(k)-}
 \, .
\end{align}
This iWeyl$_k$ reflection generates the $q_k$-character
\begin{align}
 \sT^{(k)}_{i,x}
 =
 \sY_{i,x}^{(k)-}
 +
 \fq_i^k \, 
 \frac{\sP_{i,x}^{(k)-} \tilde{\sP}_{i,q_1^{-1} x}^{(k)+}}
      {\sY^{(k)+}_{i,q_1^{-1} x}}
 \prod_{e: j \to i} \sY_{j,\mu_e q_1^{-1} x}^{(k)+}
 \prod_{e: i \to j} \sY_{j,\mu_e^{-1} x}^{(k)-}
 + \cdots
 \, .
\end{align}
This is a regular function in $x^k$, but not a polynomial, since its an elliptic function.
For example, the $q_k$-character for $A_1$ quiver is given by
\begin{align}
 \sT_x^{(k)}
 =
 \sY_x^{(k)-}
 +
 \fq_i^k \, 
 \frac{\sP^{(k)-}_x \tilde{\sP}^{(k)+}_{q_1^{-1} x}}
      {\sY^{(k)+}_{q_1^{-1} x}}
 \, .
\end{align}
This expression is the same as 5d theory \eqref{eq:qq-ch_k_A1}.

\section{Operator formalism}\label{sec:op}

We reformulate the NS$_k$ limit using the operator formalism of quiver gauge theory, which plays a central role in the construction of quiver W-algebras~\cite{Kimura:2015rgi,Kimura:2016dys,Kimura:2017hez}.
To begin with, we consider the deformation of the UV prepotential with the all the possible holomorphic operators~\cite{Marshakov:2006ii}
\begin{align}
 \msF_\text{UV}
 \ \longrightarrow \
 \msF_\text{UV}
 + \sum_{i \in \Gamma_0} \sum_{n = 1}^\infty t_{i,n} \, \bY_i^{[n]}
 \, .
\end{align}
The coupling parameter $(t_{i,n})_{i \in \Gamma, n \in \BZ_{\ge 1}}$ is called the time variables from the analogy with the integrable hierarchy.
This deformation gives rise to the {\em potential term} in the partition function
\begin{align}
 Z_i^\text{pot}(t)
 & =
 \exp
 \left(
  \sum_{n=1}^\infty t_{i,n} \, \bY_i^{[n]}
 \right)
 \, .
\end{align}
In the presence of the potential term the observable $\bY_{i}^{[n]}$ is realized as the derivative with respect to the conjugate time variable 
\begin{align}
 \bY_i^{[n]} Z_i^\text{pot}(t)
 =
 \frac{\partial}{\partial t_{i,n}} Z_i^\text{pot}(t)
 \, .
\end{align}
This leads to the identification $\bY_i^{[n]} = \partial_{i,n} =: t_{i,-n}$, so that we have the Heisenberg algebra $\bH$ with the commutation relation $[t_{i,n}, t_{j,n'}] = \delta_{ij} \delta_{n+n',0}$.
This is the operator formalism of gauge theory.
The $t$-extended partition function, which depends explicitly on the time variables, is promoted to the $Z$-state $\ket{Z}$ in the Fock space generated by $(t_{i,n})_{i \in \Gamma_0, n \in \BZ_{\ge 1}}$, due to the operator/state correspondence in CFT.

\subsection{$Z$-state}

We have several operators in the operator formalism of gauge theory.
The $t$-extended partition function is given by the $Z$-state, which is generated by the screening charge from the vacuum state
\begin{align}
 \ket{Z}
 =
 \prod_{\mathring{x} \in \CalX_0}^\succ
 \sS_{\si(\mathring{x}),\mathring{x}}
 \ket{1}
\end{align}
where the vacuum $\ket{1}$ is a constant with the time variables $(t_{i,n})$ annihilated by the corresponding derivatives $\partial_{i,n} \ket{1} = 0$, and the screening charge is defined using the screening current
\begin{align}
 \sS_{i,x} = \sum_{k \in \BZ} S_{i,q_2^k x}
 \, .
\end{align}
The node label $\si: \CalX \to \Gamma_0$ is defined as $\si(x) = i$ for $x \in \CalX_i$.
The screening current $S_{i,x}$ for 5d gauge theory has a free field realization
\begin{align}
 S_{i,x}
 = \,
 :
 \exp
 \left(
  s_{i,0} \log x + \tilde{s}_{i,0} + \sum_{n \neq 0} s_{i,n} \, x^{-n}
 \right)
 :
\end{align}
with the commutation relation
\begin{align}
 \Big[s_{i,n}, s_{j,n'}\Big]
 =
 - \frac{1}{n} \frac{1 - q_1^n}{1 - q_2^{-n}} \,
 c_{ji}^{[n]} \, \delta_{n+n',0}
 \, ,
 \label{eq:s-commutator}
\end{align}
and the zero modes
\begin{align}
 \Big[ \tilde{s}_{i,0}, s_{j,n} \Big]
 = - \beta \, c_{ij}^{[0]} \, \delta_{n,0}
 \, , \qquad
 \beta = - \frac{\log q_1}{\log q_2} = - \frac{\epsilon_1}{\epsilon_2}
 \, .
 \label{eq:beta}
\end{align}
The matrix $(c_{ij}^{[n]})$ is obtained from the $n$-th Adams operation of the {\em mass-deformed} quiver Cartan matrix
\begin{align}
 c_{ij}
 =
 c_{ij}^+ + c_{ij}^-
 =
 (1 + q^{-1}) \delta_{ij}
 - \sum_{e:i \to j} \mu_e^{-1}
 - \sum_{e:j \to i} \mu_e q^{-1}
 \, ,
 \label{eq:q-Cartan}
\end{align}
which reproduces the quiver Cartan matrix \eqref{eq:Cartan_quiver} when $n = 0$.
The first half is defined in \eqref{eq:Cartan_half1}, and the other is given by $c_{ij}^- = q^{-1} (c_{ji}^+)^\vee$.
See \cite{Kimura:2017hez} for non-simply-laced quiver construction.

From the $Z$-state, we obtain the plain (non-$t$-extended) 5d gauge theory partition function, especially yielding vector and bifundamental hypermultiplets discussed in Sec.~\ref{sec:gauge-th}, as a correlator,
\begin{align}
 Z(t = 0)
 =
 \vev{1|Z}
 =
 \bra{1}
 \prod_{x \in \mathring{\CalX}}^\succ
 \sS_{\si(x),x}
 \ket{1}
 \, .
\end{align}
In order to incorporate the fundamental hypermultiplets, we need another vertex operator, called the $\sV$-operator,
\begin{align}
 \sV_{i,x}
 = \,
 :
 \exp
 \left(
  \sum_{n \neq 0} v_{i,n} \, x^{-n}
 \right)
 :
\end{align}
with the commutation relation
\begin{align}
 \Big[
 v_{i,n}, s_{j,n'}
 \Big]
 =
 \frac{1}{n} \frac{1}{1 - q_2^{n}}
 \, \delta_{ij} \, \delta_{n+n',0}
 \, .
 \label{eq:vs-commutator}
\end{align}
The product of the screening current and the $\sV$-operator is given by
\begin{align}
 \sV_{i,x'} S_{i,x}
 =
 \left( \frac{x}{x'}; q_2 \right)_\infty^{-1}
 : \sV_{i,x'} S_{i,x} :
 \, , \qquad
 S_{i,x} \sV_{i,x'} 
 =
 \left( q_2 \frac{x'}{x}; q_2 \right)_\infty
 : \sV_{i,x'} S_{i,x} :
 \, .
\end{align}
Thus the contribution of fundamental and anti-fundamental hypermultiplets is realized using the $\sV$-operator
\begin{align}
 \ket{Z}
 =
 \left(
 \prod_{x \in \CalX_\text{f}} \sV_{\si(x),x}
 \right)
 \left(
 \prod_{\mathring{x} \in \mathring{\CalX}}^\succ
 \sS_{\si(\mathring{x}),\mathring{x}}
 \right)
 \left(
 \prod_{x \in \CalX_\text{af}} \sV_{\si(x),x}
 \right)
 \ket{1}
\end{align}
where $\CalX_\text{f} = \{\mu_{i,f}\}_{i \in \Gamma_0,f \in [1,\ldots,\sn_i^\text{f}]}$ and $\CalX_\text{af} = \{\tilde{\mu}_{i,f}\}_{i \in \Gamma_0,f \in [1,\ldots,\sn_i^\text{af}]}$ are sets of the multiplicative mass parameters.
We remark that the shift of fundamental mass $\mu_{i,f} \to \mu_{i,f} q^{-1}$ is necessary for the precise agreement with the gauge theory definition \eqref{eq:Z_5d}.
The corresponding plain partition function is given by closing the $Z$-state with the dual vacuum $\bra{1}$,
\begin{align}
 Z(t=0)
 =
 \vev{1 | Z}
 =
 \bra{1}
 \left(
 \prod_{x \in \CalX_\text{f}} \sV_{\si(x),x}
 \right)
 \left(
 \prod_{\mathring{x} \in \mathring{\CalX}}^\succ
 \sS_{\si(\mathring{x}),\mathring{x}}
 \right)
 \left(
 \prod_{x \in \CalX_\text{af}} \sV_{\si(x),x}
 \right)
 \ket{1}
 \, .
\end{align}
A similar construction is applicable for 6d gauge theory.
See \cite{Kimura:2016dys} for explicit argument.

In the NS limit $q_2 \to 1$, the commutation relations \eqref{eq:s-commutator} and \eqref{eq:vs-commutator} for all the oscillators become apparently singular.
This singularity corresponds to the diverging behavior of the partition function in the NS limit \eqref{eq:ZW}.
In the NS$_k$ limit, on the other hand, the singularity appears only for the modes $n \in k\BZ$.
This reflects the fact that the modes with $n \in k \BZ$ would be center elements in this limit as discussed in the following.

\subsection{iWeyl reflection and $qq$-character}

In the operator formalism, the $\sY$-function discussed in Sec.~\ref{sec:saddle} is promoted to the operator, which is used to construct the generating current of the quiver W-algebra.
The free field realization of the $\sY$-operator yields
\begin{align}
 \sY_{i,x}
 & =
 q_1^{\tilde{\rho}_i}
 :
 \exp
 \left(
  y_{i,0} + \sum_{n \neq 0} y_{i,n} \, x^{-n}
 \right)
 :
\end{align}
with the commutation relation
\begin{align}
 \Big[ y_{i,n}, y_{j,n'} \Big]
 =
 - \frac{1}{n} (1 - q_1^n) ( 1 - q_2^n) \, \tilde{c}_{ij}^{[-n]} \,
 \delta_{n+n',0}
 \, .
 \label{eq:y-osc}
\end{align}
Here $\tilde{\rho}_i$ is the Weyl vector $\displaystyle \sum_{j \in \Gamma_0} \tilde{c}_{ji}^{[0]}$, and the matrix $(\tilde{c}_{ij})$ is the inverse of the mass-deformed Cartan matrix \eqref{eq:q-Cartan}.
If the quiver Cartan matrix is not invertible, we deal with the $q_1$ factor separately.
Since the $y$- and $s$-oscillators obey the relation
\begin{align}
 \Big[ y_{i,n}, s_{j,n'} \Big]
 =
 - \frac{1}{n} (1 - q_1^n) \, \delta_{ij} \, \delta_{n+n',0}
 \, , \qquad
 \Big[ \tilde{s}_{i,0}, y_{j,0} \Big]
 = - \delta_{ij} \log q_1
 \, ,
\end{align}
the two $\sY$-functions are obtained as follows, up to some trivial factors,
\begin{align}
 \vev{\sY_{i,x}^+}
 =
 \bra{1} \sY_{i,x} \ket{Z}
 \, , \qquad
 \vev{\sY_{i,x}^-}
 =
 \bra{Z} \sY_{i,x} \ket{1}
 \, .
\end{align}
Thus the $\sY$-operator is the fundamental operator for the $\sY$-functions in this sense.

We then define the iWeyl reflection generator, called the $\sA$-operator,
\begin{align}
 \sA_{i,x}
 =
 q_1
 :
 \exp
 \left(
  a_{i,0} + \sum_{n \neq 0} a_{i,n} \, x^{-n}
 \right)
 :
\end{align}
which plays a role as a ``root'', while the $\sY$-operator is interpreted as a ``weight'' associated with the quiver, since they are related to each other via the quiver Cartan matrix
\begin{align}
 a_{i,n} = y_{j,n} \, c_{ji}^{[n]}
 \, .
\end{align}
Thus the $\sA$-operator has an alternative expression in terms of the $\sY$-operators
\begin{align}
 \sA_{i,x}
 = \
 :\sY_{i,x} \sY_{i,qx}
 \prod_{e:i \to j} \sY_{j,\mu_e^{-1} q x}^{-1}
 \prod_{e:j \to i} \sY_{j,\mu_e x}^{-1} 
 :
\end{align}
The $a$-oscillator obeys the commutation relation
\begin{align}
 \Big[ a_{i,n}, a_{j,n'} \Big]
 =
 - \frac{1}{n} (1 - q_1^n)(1 - q_2^n) \,
 c_{ji}^{[n]} \, \delta_{n+n',0}
 \, ,
\end{align}
and also
\begin{align}
 \Big[ y_{i,n}, a_{j,n'} \Big]
 =
 - \frac{1}{n} (1 - q_1^n)(1 - q_2^n) \, \delta_{ij} \, \delta_{n+n',0}
 \, .
\end{align}

The generating current of quiver W-algebra W($\Gamma$) is constructed as the $qq$-character of the fundamental representation associated to each node of the quiver $\Gamma$.
Starting with the operator $\sY_{i,x}$, as a highest weight, it is obtained by the iWeyl reflection generated by the ``root'' operator $\sA_{i,x}$,
\begin{align}
 \sT_{i,x}
 =
 \sY_{i,x}
 + \, : \sY_{i,x} \sA_{i,q^{-1}x}^{-1} :
 + \cdots
 \, .
 \label{eq:qq-ch_op}
\end{align}
We can show that the operator version of the $qq$-character defined here commutes with the screening charge~\cite{Kimura:2015rgi}
\begin{align}
 \Big[ \sT_{i,x}, \sS_{j,x'} \Big] = 0
 \, .
\end{align}
This property assures that the $qq$-character is a holomorphic generating current of quiver W-algebra.
If involving a product of $\sY$-operators, we need to multiply the $\msS$-factor \eqref{eq:S-factor}
\begin{align}
 :\prod_{l=1}^{n} \sY_{i,x_l}:
 \ \longrightarrow \
 \sum_{I \cup J = \{1,\ldots,n\}}
 \left(
 \prod_{l \in I, m \in J}
 \msS \left( \frac{x_l}{x_m} \right)
 \right)
 :
 \left( \prod_{l=1}^{n} \sY_{i,x_l} \right)
 \left( \prod_{m \in J} \sA_{i,q^{-1} x_m}^{-1} \right)
 :
 \, .
 \label{eq:ref_prod}
\end{align}
For example, for $n=2$, the iWeyl reflection is given by
\begin{align}
 :\sY_{i,x} \sY_{i,x'}:
 + \,
 \msS\left( \frac{x'}{x} \right)
 :\sY_{i,x} \sY_{i,x'} \sA_{i,q^{-1}x}^{-1}:
 + \,
 \msS\left( \frac{x}{x'} \right)
 :\sY_{i,x} \sY_{i,x'} \sA_{i,q^{-1}x'}^{-1}:
 +
 :\sY_{i,x} \sY_{i,x'} \sA_{i,q^{-1}x}^{-1} \sA_{i,q^{-1}x'}^{-1}:
 \, .
\end{align}
In the collision limit $x' \to x$, we have to consider the derivative expansion~\cite{Kimura:2015rgi}.

The action of the iWeyl reflection closes for finite-type quiver such that the determinant of the corresponding quiver Cartan is positive: $\det c > 0$.
For example, the $qq$-character for $A_1$ quiver, which consists of a single node, is given by
\begin{align}
 \sT_{1,x}
 =
 \sY_{1,x} + \sY_{1,q^{-1} x}^{-1}
 \, ,
\end{align}
corresponding to the fundamental (2-dimensional) representation of $SU(2)$ since the $\sA$-operator is given by
\begin{align}
 \sA_{1,x} = \sY_{1,x} \sY_{1,q x}
 \, .
\end{align}
The operator $\sT_{1,x}$ plays a role as a generating current of the $q$-deformed Virasoro algebra W($A_1$).
Even if $\det c \le 0$, we can similarly consider the $qq$-character, as a holomorphic generating current of W($\Gamma$) algebra, which consists of infinite monomials of $\sY$-operators.
Taking the classical limit $q_{1,2} \to 1$, the $qq$-character reproduces the Seiberg--Witten geometry for $\Gamma$-quiver gauge theory~\cite{Nekrasov:2012xe}.

\subsection{NS and NS$_k$ limit}

We consider the NS limit of the operator formalism.
The commutators of the $y$- and $a$-oscillators, e.g. \eqref{eq:y-osc}, have the factor $(1 - q_2^n)$ for $n \in \BZ$.
This means that all the $\sY$ and $\sA$ operators become commutative in the NS limit $q_2 \to 1$, and thus we obtain the corresponding commuting $\sT$-operator, interpreted as the commutative transfer matrix of the quantum integrable system associated to the quiver $\Gamma$, and also as the $q$-character of quantum affine algebra $U_\hbar(\widehat{\g}_\Gamma)$~\cite{Frenkel:1998} reduced from the $qq$-character~\cite{Nekrasov:2013xda}.
Here $\g_\Gamma$ is the Lie algebra associated to the quiver $\Gamma$ if it is of finite-type.
In terms of $q$-deformed W-algebra, it is reduced to the classical commutative Poisson algebra, which is isomorphic to the center of quantum affine algebra at the classical level and also the representation ring described by the $q$-character, as a consequence of the geometric $q$-Langlands correspondence.

In the NS$_k$ limit, there appears a similar structure.
It has been shown by Bouwknegt--Pilch~\cite{Bouwknegt1997qru,Bouwknegt:1998da} that the $q$-deformed W-algebra gives rise to the extended center when either $q_{1,2}$ is taken to be the root of unity, which is indeed the NS$_k$ limit.
We would provide gauge theory proof for their statement.

The commutation relations for the $s$- and $v$-oscillators, e.g.~\eqref{eq:s-commutator}, have the factor $(1-q_2^n)$ in the denominator, which become singular for $n \in k \BZ$ in the NS$_k$ limit.
Since these oscillators are used to construct the $Z$-state, namely the gauge theory partition function, such a singularity corresponds to the diverging behavior of the partition function in this limit, as shown in \eqref{eq:ZW_5d}.
The other modes $n \not \in k \BZ$ are regular, but do not contribute to the diverging asymptotics of the partition function.

The commutation relation for the $y$- and $a$-oscillators, e.g.~\eqref{eq:y-osc}, have the factor $(1-q_2^n)$ in the numerator, on the other hand.
Therefore the modes $n \in k\BZ$ become commutative, while the others $n \not \in k\BZ$ still obey a nontrivial relation.
This implies that we can extract the central element by focusing on specific part of the oscillators.
For this purpose, we define the $\sY^{(k)}$-operator, analogous to \eqref{eq:Yk-functions},
\begin{align}
 \sY_{i,x}^{(k)}
 = \
 : \prod_{r=0}^{k-1} \sY_{i,\zeta_k^r x}: \
 =
 q_1^{k\tilde{\rho}_i}
 \exp 
 \left(
 k \sum_{n \in k \BZ} y_{i,n} \, x^{-n}
 \right)
 \, ,
 \label{eq:Yk-op}
\end{align}
which contains only the commutative modes $n \in k\BZ$ in the NS$_k$ limit.
Similarly we define the $\sA^{(k)}$-operator
\begin{align}
 \sA_{i,x}^{(k)}
 = \
 : \prod_{r=0}^{k-1} \sA_{i,\zeta_k^r x}: \
 =
 q_1^{k}
 \exp 
 \left(
 k \sum_{n \in k \BZ} a_{i,n} \, x^{-n}
 \right)
 \, .
 \label{eq:Ak-op}
\end{align}
We then construct the degree-$k$ $qq$-character, which we call the $qq_k$-character, using these operators
\begin{align}
 \sT_{i,x}^{(k)}
 =
 \sY_{i,x}^{(k)}
 \ +
 : \sY_{i,x}^{(k)} \sA_{i,q^{-1} x}^{(k)-1} :
 + \cdots
 \label{eq:q-ch_op}
\end{align}
which reproduces the result from the gauge theory analysis in the NS$_k$ limit \eqref{eq:q-ch_k}, namely the $q_k$-character, by taking the average with respect to the $Z$-state.
As long as the quiver is fixed, we can use the operator version of the $qq_k$-character \eqref{eq:q-ch_op} with arbitrary matter content.
In this sense, the operator formalism provides a universal prescription to construct the $qq$-character in gauge theory.

Apparently the $q_k$-character generates the central elements of quiver W-algebra in the NS$_k$ limit, since it is constructed only with the commutative oscillators $n \in k \BZ$ in the limit.
We remark that the $qq_k$-character \eqref{eq:q-ch_op} does not commute with the screening charge for generic $(q_1,q_2)$.%
\footnote{%
We can construct the screening charge which commutes with the $qq_k$-character by using the degree-$k$ screening current (and also the $\sV^{(k)}$-operator),
\begin{align}
 S_{i,x}^{(k)} =
 \ : \prod_{r=0}^{k-1} S_{i,\zeta_k^r x} :
 \, , \qquad
 \sV_{i,x}^{(k)} =
 \ : \prod_{r=0}^{k-1} \sV_{i,\zeta_k^r x} :
 \, .
\end{align}
These operator define $\BZ_k$-invariant gauge theory partition function under $(q_1, q_2) \to (\zeta_k^m q_1, \zeta_k^n q_2)$ for arbitrary integers $(m,n)$, since the partition function depends only on $(q_1^k, q_2^k)$ in this case.
}
However, after taking the NS$_k$ limit, the $q_k$-character is constructed as a higher representation $q$-character, namely a product of the ordinary $q$-characters.
Following Bouwknegt--Pilch~\cite{Bouwknegt1997qru,Bouwknegt:1998da} (but through another path), we show that the $q_k$-character defined as the NS$_k$ reduction of the $qq_k$-character \eqref{eq:q-ch_op} is given by
\begin{align}
 \lim_{q_2 \to \zeta_k}
 \sT_{i,x}^{(k)}
 =
 \lim_{q_2 \to \zeta_k} 
 \left(
  \prod_{0 \le r<s \le k-1} f_{ii} \left( \zeta_k^{r-s} \right)
 \right)
 : \prod_{r=0}^{k-1} \sT_{i,\zeta_k^r x} :
 \label{eq:qk-ch_BP}
\end{align}
where we define
\begin{align}
 f_{ij}(x)
 =
 \exp
 \left(
 \sum_{n=1}^\infty (1 - q_1^n)(1 - q_2^n) \,
 \tilde{c}_{ij}^{[-n]} \, x^n
 \right)
\end{align}
which appears in the OPE of $\sY$-operators
\begin{align}
 \sY_{i,x} \sY_{j,x'}
 =
 f_{ij} \left( \frac{x'}{x} \right)^{-1}
 :\sY_{i,x} \sY_{j,x'}:
 \, .
 \label{eq:YY-OPE}
\end{align}
Since each $qq$-character $\sT_{i,\zeta_p^r x}$ contains $\sY_{i,\zeta_p^r x}$ as a highest weight, as defined in \eqref{eq:qq-ch_op}, and the $\sY$-operator obeys the OPE \eqref{eq:YY-OPE}, the highest term is given by the $\sY^{(k)}$-operator,
\begin{align}
 \left(
 \prod_{0 \le r<s \le k-1} f_{ii} \left( \zeta_k^{r-s} \right)
 \right)
 : \prod_{r=0}^{k-1} \sT_{i,\zeta_k^r x} : \
 =
 \sY_{i,x}^{(k)} + \cdots
 \, .
\end{align}
This expression holds for generic $q_{1,2}$.
Then we consider the subleading terms generated by the iWeyl reflection.
We apply the formula \eqref{eq:ref_prod} to the highest $\sY^{(k)}$-operator, since it is a product of $\sY$-operators \eqref{eq:Yk-op},
\begin{align}
 \sY_{i,x}^{(k)}
 \ \longrightarrow \
 \sum_{I \cup J = \{0,\ldots,k-1\}}
 \left(
 \prod_{l \in I, m \in J}
 \msS \left( \zeta_k^{l-m} \right)
 \right)
 :
 \left( \prod_{l=0}^{k-1} \sY_{i,x_l} \right)
 \left( \prod_{m \in J} \sA_{i,q^{-1} x_m}^{-1} \right)
 :
 \, . 
\end{align}
In the NS$_k$ limit, the product of $\msS$-factors becomes zero unless $I = \emptyset$ or $J = \emptyset$ because
\begin{align}
 \lim_{q_2 \to \zeta_k} \msS(\zeta_k^{-1}) = 0
 \, .
\end{align}
Thus, in the NS$_k$ limit, only the configuration with $I = \emptyset$, $J = \{0,\ldots, k-1\}$ contributes to the reflection, which corresponds to the reflection generated by the $\sA^{(k)}$-operator \eqref{eq:Ak-op}.
This proves the relation \eqref{eq:qk-ch_BP}.

\section{Discussion}\label{sec:discussion}

In this paper we have discussed the twisted Nekrasov--Shatashvili limit $q_2 \rightarrow \zeta_k$, which we call the NS$_k$ limit, of 5d (K-theoretic) and 6d (elliptic) quiver gauge theory.
In this limit, we have found a diverging behavior, and obtained the saddle point equation for the gauge theory partition function similar to the ordinary NS limit $(k=1)$.
In particular, the NS$_k$ limit yields an extra symmetry under the $\BZ_k$ transformation for the multiplicative parameter $x \to \zeta_k x$.
Indeed only the $\BZ_k$ symmetric sector contributes to the asymptotic behavior of the partition function, which is interpreted as the twisted superpotential for the effective 2d gauge theory, as proposed in \cite{Nekrasov:2009rc} for the ordinary NS limit.
Although the root of unity limit of the equivariant parameter plays a role as the orbifold projection~\cite{Kimura:2011zf,Kimura:2011gq}, it is not the case for the current situation since we keep the other equivariant parameter finite $q_1 \neq 1$.
We still have a finite compactification circle $S^1$, and the corresponding geometry should not be $\BC \times \left(\BC/\BZ_k\right)$. 
Due to this $\BZ_k$ symmetry, we only have the $\BZ_k$ invariant gauge theory observables, whose expectation value parametrizes the Coulomb branch of the SUSY vacua.

The result obtained in this paper also sheds a new light on a connection between W-algebra and quantum affine algebra.
The geometric $q$-Langlands correspondence is an isomorphism between the conformal blocks of the classical $q$-deformed W-algebra and quantum affine algebra at the critical level.
Then its quantization is the correspondence between W-algebra and quantum affine algebra away from the classical and the critical limit, respectively~\cite{Aganagic:2017smx}.
From this point of view, since $\Gamma$-quiver gauge theory with generic equivariant parameter $(q_1, q_2)$ corresponds to the full W-algebra W$_{q_1,q_2}(\Gamma)$~\cite{Kimura:2015rgi,Kimura:2016dys,Kimura:2017hez}, the ordinary NS limit is a reduction to the classical commutative Poisson algebra, which is isomorphic to quantum affine algebra at the critical level~\cite{Nekrasov:2013xda}.
More explicitly, the parameters are related to each other:
\begin{align}
 q_1 = \hbar^{\beta(\sk + h^\vee)}
 \, , \qquad
 q_2 = \hbar^{-(\sk + h^\vee)}
 \, .
 \label{eq:Langlands}
\end{align}
where $\beta$ is defined in \eqref{eq:beta}, and the level $\sk$ and the dual Coxeter number $h^\vee$ of quantum affine algebra $U_\hbar(\widehat{\g}_\Gamma)$ with the deformation parameter $\hbar$.%
\footnote{%
We here consider the simply-laced algebra ${^{L}\g = \g}$ with the lacing number $m=1$ for simplicity.
}
Thus the NS$_{k=1}$ limit $q_1 \neq 1$, $q_2 \to 1$ means the classical limit and the critical level
\begin{align}
 \beta = \infty
 \, , \qquad
 \sk = - h^\vee
 \, .
 \label{eq:cl_lim}
\end{align}
Applying the parametrization \eqref{eq:Langlands} to the NS$_k$ limit $q_2 \to \zeta_k$, on the other hand, we obtain
\begin{align}
 q_1 = \exp \left( - \frac{2\pi \iota}{k} \beta \right)
 \, , \quad
 q_2 = \exp \left( \frac{2\pi \iota}{k} \right)
 \, , \quad
 \hbar
 =
 \exp \left( - \frac{2 \pi \iota}{k (\sk + h^\vee)} \right)
 \, .
\end{align}
We remark that the deformation parameter becomes $\hbar \stackrel{k \to 1}{\longrightarrow} \exp \left( - 2\pi \iota / (\sk + h^\vee) \right)$, which is associated with the coupling constant of $\g$-Chern--Simons theory at level $\sk$.
Although, as shown in this paper, the NS$_k$ limit plays a quite similar role to the ordinary NS limit, we don't need to take the classical limit \eqref{eq:cl_lim}: we can keep the parameters $\beta$ and $\sk$ generic.
This manifests that there is no counterpart of the NS$_k$ limit in 4d gauge theory.

\appendix

\section{Asymptotic formulae}
\label{sec:asymptotic}

We summarize the formulae used to study the saddle point equation of the gauge theory partition function in the NS and NS$_k$ limit.

\subsection{5d theory}
\label{sec:asymptotic_5d}

\subsubsection{NS limit}

Let $q = e^\epsilon$, then the quantum dilogarithm function has the $\epsilon$-expansion
\begin{align}
 (z;q)_\infty^{-1}
 & =
 \exp
 \left(
 \sum_{m=1}^\infty \frac{z^m}{m(1 - q^m)}
 \right)
 =
 \exp
 \left(
  - \frac{1}{\epsilon} \Li_2(z) + O(\epsilon^0)
 \right)
 \, .
\end{align}
The subleading terms have an explicit expression in terms of the Bernoulli number.
We take the following combination of the quantum dilogs which appears in the gauge theory partition function
\begin{align}
 \left( q z; q_2 \right)_\infty^{-1}
 \left( q_2 z; q_2 \right)_\infty
 & \stackrel{\epsilon_2 \to 0}{\longrightarrow} \
 \exp
 \left(
  - \frac{1}{\epsilon_2} \left( \Li_2(q_1 z) - \Li_2(z) \right) 
 \right)
 \nonumber \\
 & =:
 \exp \left( - \frac{1}{\epsilon_2} L(z;q_1) \right)
\end{align}
where we define
\begin{align}
 L(z;q_1) = \Li_2(q_1 z) - \Li_2(z)
 \label{eq:L-func}
\end{align}
and the derivative yields
\begin{align}
 \frac{d}{d \log z} L(z; q_1)
 & =
 - \log (1 - q_1 z) + \log (1 - z)
 = \log \left( \frac{1 - z}{1 - q_1 z} \right)
 \, .
\end{align}

\subsubsection{NS$_k$ limit}

We similarly consider the NS$_k$ limit $q_2 = \zeta_k e^{\epsilon}$ with $\epsilon \to 0$ where the $k$-th root of unity defined $\zeta_k = \exp \left( 2 \pi \iota / k \right)$.
Let $q = e^\epsilon$ again, and it yields
\begin{align}
 (z; \zeta_k q)_\infty^{-1}
 & =
 \exp
 \left( - \frac{1}{k^2 \epsilon} \Li_2(z^k) + O(\epsilon^0) \right)
 \, ,
\end{align}
where we used
\begin{align}
 \frac{1}{1 - (\zeta_k q)^m}
 = \sum_{n = 0}^\infty
 \left( q^{kmn} + \zeta_k q^{kmn+m} + \cdots \zeta_k^{k-1} q^{kmn+(k-1)m} \right)
 = \sum_{r=0}^{k-1} \frac{\zeta_k^{r} q^{mr}}{1 - q^{km}}
\end{align}
and
\begin{align}
 \sum_{r=0}^{k-1} \Li_n(\zeta_k^r z)
 = \sum_{r=0}^{k-1} \sum_{m=1}^\infty \frac{\zeta_k^{rm} z^{m}}{m^n}
 = \frac{1}{k^{n-1}} \Li_n(z^k)
 \, .
\end{align}
Then, in the NS$_k$ limit $q_2 = \zeta_k e^{\epsilon_2}$ with $\epsilon_2 \to 0$, we have
\begin{align}
 \left( \zeta_k q_1 e^{\epsilon_2} z; \zeta_k e^{\epsilon_2} \right)_\infty^{-1}
 \left( \zeta_k e^{\epsilon_2} z; \zeta_k e^{\epsilon_2} \right)_\infty
 & \stackrel{\epsilon_2 \to 0}{\longrightarrow} \
 \exp
 \left(
  - \frac{1}{k^2 \epsilon_2}
 \left(
  \Li_2(q_1^k z^k) - \Li_2(z^k)
 \right)
 \right)
 \nonumber \\
 & =:
 \exp
 \left(
  - \frac{1}{k \epsilon_2} L_k(z;q_1)
 \right)
\end{align}
where we define
\begin{align}
 L_k(z;q_1)
 & =
 \frac{1}{k} \left( \Li_2(q_1^k z^k) - \Li_2(z^k) \right)
 \label{eq:Lk-func} 
\end{align}
and
\begin{align}
 \frac{d}{d \log z} L_k(z;q_1)
 & =
 - \log (1 - q_1^k z^k) + \log (1 - z^k)
 = \log \left( \frac{1 - z^k}{1 - q_1^k z^k} \right)
 \, .
\end{align}

\subsection{6d theory}
\label{sec:asymptotic_6d}

\subsubsection{NS limit}

We begin with the expansion of the elliptic gamma function
\begin{align}
 \Gamma(z;p,q)
 =
 \prod_{n,m \ge 0}
 \frac{1 - z^{-1} p^{n+1} q^{n+1}}{1 - z p^n q^n}
 =
 \exp
 \left(
  \sum_{m \neq 0} \frac{z^m}{m(1 - p^m)(1 - q^m)}
 \right)
 \, .
\end{align}
Putting $q = e^\epsilon$, the asymptotic behavior in the limit $\epsilon \to 0$ is given by
\begin{align}
 \Gamma(z;p,q)
 & =
 \exp
 \left(
  - \frac{1}{\epsilon} \sum_{m \neq 0} \frac{z^m}{m^2(1 - p^m)}
  + O(\epsilon^0)
 \right)
 \nonumber \\
 & =
 \exp
 \left(
  - \frac{1}{\epsilon} \Li_2(z;p)
  + O(\epsilon^0)
 \right)
 \, ,
\end{align}
where we define an elliptic analogue of the polylogarithm
\begin{align}
 \Li_k(z;p) = \sum_{m \neq 0} \frac{z^m}{m^k(1-p^m)}
 \, .
\end{align}
This is reduced to the ordinary polylogarithm in the limit $p \to 0$ ($\Im \tau \to \infty$), and obeys the similar descendant relation
\begin{align}
 \frac{d}{d \log z} \Li_k(z;p) = \Li_{k-1}(z;p)
 \, .
\end{align}
In particular, we have
\begin{align}
 \Li_{1}(z;p)
 = \sum_{m \neq 0} \frac{z^m}{m(1-p^m)}
 = - \log \theta(z;p)
 \, .
\end{align}
Then the elliptic gamma function ratio used in the 6d gauge theory partition function has the asymptotic behavior
\begin{align}
 \frac{\Gamma(qz; p, q_2)}{\Gamma(q_2 z; p, q_2)}
 & \stackrel{\epsilon_2 \to 0}{\longrightarrow} \
 \exp
 \left(
  - \frac{1}{\epsilon_2} \left( \Li_2 (q_1 z;p) - \Li_2(z;p) \right)
 \right)
 \nonumber \\
 & =:
 \exp
 \left(
  - \frac{1}{\epsilon_2} L(z;q_1;p)
 \right)
\end{align}
with the elliptic $L$-function defined
\begin{align}
 L(z;q_1;p)
 =
 \Li_2 (q_1 z;p) - \Li_2(z;p)
 \, ,
\end{align}
and its derivative yields
\begin{align}
 \frac{d}{d \log z} L(z;q_1;p)
 = \log
 \left(
  \frac{\theta(z;p)}{\theta(q_1 z;p)}
 \right)
 \, .
\end{align}

\subsubsection{NS$_k$ limit}

We consider the elliptic NS$_k$ limit.
Put $q_2 = \zeta_k e^{\epsilon}$ with $\epsilon \to 0$ and the $k$-th root of unity $\zeta_k = \exp \left( 2 \pi \iota / k \right)$ as before.
The elliptic gamma function behaves in the NS$_k$ limit as
\begin{align}
 \Gamma(z;p,\zeta_k q)
 =
 \exp
 \left( - \frac{1}{k^2 \epsilon} \Li_2(z^k;p^k) + O(\epsilon^0) \right)
\end{align}
since
\begin{align}
 \sum_{r=0}^{k-1} \Li_n(\zeta_k^r z;p)
 = \sum_{r=0}^{k-1} \sum_{m \neq 0} \frac{\zeta_k^{rm} z^{m}}{m^n(1-p^m)}
 = \frac{1}{k^{n-1}} \Li_n(z^k;p^k)
 \, .
\end{align}
Then the elliptic gamma function ratio is given by
\begin{align}
 \frac{\Gamma(q z; p, q_2)}{\Gamma(q_2 z; p, q_2)}
 \ \stackrel{q_2 \to \zeta_k e^{\epsilon_2}}{\longrightarrow} \
 \frac{\Gamma(\zeta_k q_1 e^{\epsilon_2} z; p, \zeta_k e^{\epsilon_2})}
      {\Gamma(\zeta_k e^{\epsilon_2} z; p, \zeta_k e^{\epsilon_2})}
 \ \stackrel{\epsilon_2 \to 0}{\longrightarrow} \
 \exp
 \left( - \frac{1}{k\epsilon_2} L_k(z;q_1;p) \right)
\end{align}
where we define
\begin{align}
 L_k(z;q_1;p)
 =
 \Li_2 (q_1^k z^k;p^k) - \Li_2(z^k;p^k)
 \, . \label{eq:Lkp-func}
\end{align}
with the derivative
\begin{align}
 \frac{d}{d \log z}
 L_k(z;q_1;p)
 =
 \log
 \left(
 \frac{\theta(z^k;p^k)}{\theta(q_1^k z^k;p^k)}
 \right)
 \, .
\end{align}


\bibliographystyle{../wquiver/utphysurl}
\bibliography{../wquiver/wquiver} 

\end{document}